\documentclass[aps, onecolumn, pra, floatfix, superscriptaddress, 10pt]{revtex4-2}
\usepackage{graphicx} 
\usepackage{subcaption} 
\usepackage{pgfplots,float,pgfplotstable} 
\pgfplotsset{compat=1.15}

\usepackage{csvsimple} 
\usepackage{bm}
\usepackage{bbold}
\usepackage{physics,amsmath,amssymb,amsfonts,amsthm}
\usepackage{ulem}

\usepackage{hyperref}
\hypersetup{
	colorlinks,    
	linkcolor=blue,     
	citecolor=blue,   
	urlcolor=blue    
}	

\usepackage{listings}
\newfloat{Figure}{htbp}{figs}
\graphicspath{{./Figures/}}

\pgfplotstableset{
	my num iterating settings/.style={%
		/pgfplots/table/display columns/#1/.style={
			postproc cell content/.style={
				@cell content/.add={\small}{}
			}
		}
	}
}

\usepackage{array, multirow, makecell} 
\newcolumntype{R}[1]{>{\raggedleft\arraybackslash }b{#1}}
\newcolumntype{L}[1]{>{\raggedright\arraybackslash }b{#1}}
\newcolumntype{C}[1]{>{\centering\arraybackslash }b{#1}}

\providecommand{\ID}{\ensuremath{\mathbb{1}}} 

\begin{document}
	
	\title{Topological quantum compilation of two-qubit gates}
	
	\author{Phillip C. Burke}
	\affiliation{School of Physics, University College Dublin, Belfield, Dublin 4, Ireland}
	\affiliation{Centre for Quantum Engineering, Science, and Technology, University College Dublin, Dublin 4, Ireland}
	\affiliation{Department of Theoretical Physics, Maynooth University, Maynooth, Kildare, Ireland}
	
	\author{Christos Aravanis}
	\affiliation{The University of Sheffield International College, United Kingdom}
	\affiliation{Department of Computer Science,
		Czech Technical University in Prague, Czech Republic}
	
	\author{Johannes Aspman}
	\affiliation{Department of Computer Science,
		Czech Technical University in Prague, Czech Republic}
	
	\author{Jakub Mare\v{c}ek}
	\affiliation{Department of Computer Science,
		Czech Technical University in Prague, Czech Republic}
	
	\author{Ji\v{r}\'i Vala}
	\affiliation{Department of Theoretical Physics, Maynooth University, Maynooth, Kildare, Ireland}
	
	\date{\today}

	\begin{abstract}
		
		We investigate the topological quantum compilation of two-qubit operations within a system of Fibonacci anyons. Our primary goal is to generate gates that are approximately leakage-free and equivalent to the controlled-NOT (CNOT) gate up to single-qubit operations. These gates belong to the local equivalence class [CNOT]. Additionally, we explore which local equivalence classes of two-qubit operations can be naturally generated by braiding Fibonacci anyons. We discovered that most of the generated classes are located near the edges of the Weyl chamber representation of two-qubit gates, specifically between the local equivalence classes of the identity $[\ID]$ and [CNOT], and between those of the double-controlled-NOT [DCNOT] and [SWAP]. Furthermore, we found a numerically exact implementation of a local equivalent of the SWAP gate using a sequence of only nine elements from the Fibonacci braiding gate set.
	\end{abstract}
	
	\maketitle
	
	\section{Introduction}
	
	Quantum computation \cite{Nielsen_00} can solve an important class of computational problems with a considerable speed-up compared to its classical counterpart. Nevertheless, it faces important challenges due to quantum errors in the computation process. In order for quantum computation to achieve its full quantum power, it has to go beyond the intrinsic limitations of noisy intermediate-scale quantum devices \cite{Preskill_18}. This requires a complex fault-tolerant quantum computing architecture \cite{Shor_96} which is based on quantum error correction with strict error-correction thresholds \cite{Knill_98, Kitaev_03, Aharonov_08}. 
	
	Topological quantum computation \cite{Freedman_03, Nayak_08, Wang_10, Pachos_12} promises to bypass these requirements by storing and processing quantum information in topological quantum states. These states are effectively described by topological quantum field theory from which they inherit robustness against errors, for example, due to undesired interactions with stray particles or physical fields. When realized, for example, in a two-dimensional many-body quantum system, these states correspond to ground states with a finite degree of degeneracy, which depends on topological properties of the underlying two-dimensional manifold such as its genus -- the number of defects such as punctures or the number of quasi-particle excitations. 
	
	Point-like excitations, specifically in two-dimensional topological systems, are called anyons as they are characterized by fractional quantum statistics which are linked to irreducible representations of the braid group \cite{Kassel_08}. Braiding anyons results in a transformation of the quantum state of the anyonic system which can be Abelian or non-Abelian. In the non-Abelian case, this transformation corresponds to a unitary rotation of a state vector within the manifold of degenerate ground states of the topological many-body system. Braiding of anyons thus implements quantum computing operations on $n$ quantum bits or qubits, encoded in the ground-state manifold with the Hilbert space dimension $d \ge 2^n$. An important attribute of anyon theory from the point of view of quantum computation is universality, that is, whether braiding permits universal quantum computing operations from the unitary group $U(2^n)$.
	
	Fibonacci anyons, which are at the center of our interest, are known to be universal for quantum computation \cite{Freedman_02}. The Fibonacci theory \cite{Nayak_08} consists of two topological charges, or $q$-spins, the trivial charge $\ID$ and the Fibonacci anyon charge $\tau$, which satisfy the single non-trivial fusion rule
	\begin{equation}
		\ID \times \tau = \ID + \tau,
	\end{equation}
	which makes this theory particularly simple. The fusion rules allow us to define an orthogonal basis for the system of two Fibonacci anyons using a convenient notation introduced by Bonesteel et al. \cite{Bonesteel_BraidTQC_PRL2005}. The state $\ket{(\bullet,\bullet)_{\ID}}$ is obtained when two topological charges $\tau$ fuse to a trivial charge, and the state $\ket{(\bullet,\bullet)_{\tau}}$ when they fuse to the charge $\tau$. 
	
	In addition, Fibonacci theory is characterized by the $R$ and $F$ matrices. The $R$ matrix gives the phase shifts associated with the interchange of two anyons and is, in the basis given above, characterized by the $2\times2$ diagonal matrix
	\begin{equation}
		R = \operatorname{diag}(R^{\tau\tau}_{1},R^{\tau \tau}_{\tau}) = 
		\begin{pmatrix}
			e^{-\frac{4\pi i}{5}} & 0 \\
			0 & e^{\frac{3\pi i}{5}}
		\end{pmatrix}.
		\label{eq_Rmat}
	\end{equation}
	
	The $F$ matrix corresponds to a change of basis or re-coupling in a system of three anyons and is defined via the following relation
	\begin{equation}
		\ket{(\bullet,(\bullet,\bullet)_i)_k} = \sum_j \left[F_k^{\tau\tau\tau}\right]_{ij}~\ket{((\bullet,\bullet)_j,\bullet)_k} .
	\end{equation}
	If these anyons fuse to the trivial charge $\ID$, $F_{\ID}^{\tau\tau\tau}$ is trivially unity. Nevertheless the matrix $\left[F_\tau^{\tau\tau\tau}\right]$ is nontrivial, 
	\begin{equation}
		F = 
		\begin{pmatrix}
			F_{\ID\ID} & F_{\ID\tau}  \\
			F_{\tau\ID} & F_{\tau\tau} 
		\end{pmatrix}
		=
		\begin{pmatrix}
			\phi^{-1} & \sqrt{\phi^{-1}} \\
			\sqrt{\phi^{-1}} & \phi^{-1}
		\end{pmatrix} 
		\label{eq_Fmat}
	\end{equation}
	where $\phi = (\sqrt{5} + 1)/2$ is the golden ratio.
	
	The seminal works on topological quantum compilation \cite{Bonesteel_BraidTQC_PRL2005, Simon_06, Hormozi_TQC_PRB2007} primarily use a three-anyon encoding of a logical qubit to realize both single-qubit and two-qubit operations via braiding or weaving. In more recent work, Carnahan et al. \cite{Carnahan_16} develop a systematic approach to the generation of two-qubit anyon braids based on an iterative procedure, which was proposed earlier by Reichardt \cite{Reichardt_12}. Kliuchnikov et al. \cite{Kliuchnikov_14} developed an asymptotically optimal probabilistic polynomial algorithm to generate single qubit gates exactly and also to produce depth-optimal approximations of two-qubit gates in the Fibonacci anyon model. Relevant to the topological quantum compilation problem is also the work on topological hashing by Burrello et al. \cite{Burrello_10} and a recent application of reinforcement learning \cite{Zhang_20}. 
	
	Implementation of two-qubit gates in topological quantum computing requires elementary braiding operations acting on a Hilbert space of a larger dimension than that of two qubits. The problem of leakage in relation to topological quantum compilation, particularly in the context of the implementation of two-qubit gates, was recognized early on by Bonesteel et al. \cite{Bonesteel_BraidTQC_PRL2005}. This problem was studied later by Xu and Wan \cite{Xu_08} and later by Ainsworth and Slingerland \cite{Ainsworth_11} who focused on the leakage problem in connection with topological qubit design and showed that the requirement of leakage-free design is too restrictive and can only be realized in non-universal Ising-like anyon models. A large family of leakage-free Fibonacci braiding gates were studied by Cui et al. \cite{Cui_2Qubit_JoPA2019}, who proved that they were all non-entangling. Moreover, brute-force numerical searches failed to identify any leakage-free implementations of entangling operations using braids with a word length of up to seven.
	
	In this work, we focus on the approximate implementation of two-qubit operations in the Fibonacci anyon framework. We employ the geometric theory of two-qubit operations, developed by Zhang et al. \cite{ZhangValaBirgitta_GeometricTheory_PRA2003} in the context of quantum control. The geometric theory stems from the Cartan decomposition of the Lie group $SU(4)$ and the Makhlin local invariants \cite{Makhlin_Invariants_QIP2002} and provides a representation of two-qubit operations in terms of their local equivalence classes. In our context, a local equivalence class is defined as the set of two-qubit gates that are equivalent up to single-qubit transformations, and thus correspond to the co-set of $SU(4)$ by the subgroup $SU(2) \otimes SU(2)$. 
	
	The local invariants were previously used by DiVincenzo et al. \cite{DiVincenzo_00} to find a local equivalent of the CNOT gate in a system with the Heisenberg Hamiltonian in the context of encoded universality. M\"{u}ller et al. \cite{Muller_11} have combined optimal quantum control with the geometric theory of two-qubit operations to derive an optimization algorithm that determines the best two-qubit gate entangling in the system of trapped polar molecules and neutral atoms. This work was later followed by a generalization of the optimal control objective to include an arbitrary perfect entangler \cite{Watts_15, Goerz_15}. In all these cases, the use of geometric theory and the Makhlin local invariants in particular led to significant enlargement of the optimization target, its dimension, and size, which considerably improved both the accuracy and convergence of the optimization process. 
	
	The structure of this paper is as follows. In Sec.~\ref{sec_encoding}, we present an encoding of a two-qubit system in the Fibonacci anyons and the discrete set of gates that correspond to elementary braids and which constitute the gate alphabet. We then attempt to generate the $\operatorname{CNOT}$ operation exactly in Sec.~\ref{sec_exactCNOT}. We shift our focus to the generation of the local equivalence class $[\operatorname{CNOT}]$ in Sec.~\ref{sec_class}, wherein we present the main results. The question of what local equivalence classes are generated naturally in our system of Fibonacci anyons is addressed in Sec.~\ref{sec_natural}. We specifically show that most of the local equivalence classes of two-qubit operations which are naturally generated by Fibonacci anyon braids are located close to the edges of the Weyl chamber representation of two-qubit operations which connects the identity local equivalence classes $[\ID]$ and $[\operatorname{CNOT}]$, and the classes of the double-CNOT $[\operatorname{DCNOT}]$ and $[\operatorname{SWAP}]$. In addition, we found a numerically exact implementation of a local equivalent of SWAP with a sequence of only nine elements from the Fibonacci braiding gate alphabet. We conclude in Sec.~\ref{sec_conclusions}.
	
	
	\section{Two qubits encoding}\label{sec_encoding}
	
	\begin{figure}[htb]
		\centering
		\begin{subfigure}{0.45\linewidth}
			\centering
			\includegraphics[width=1\linewidth]{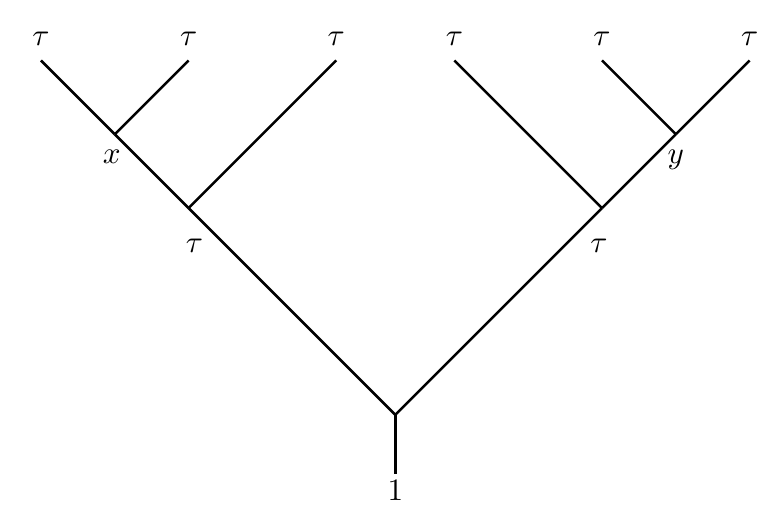}
			\caption{$\ket{xy}$ states in computational basis}
		\end{subfigure}
		\hspace*{0.5cm}
		\begin{subfigure}{0.45\linewidth}
			\centering
			\includegraphics[width=1\linewidth]{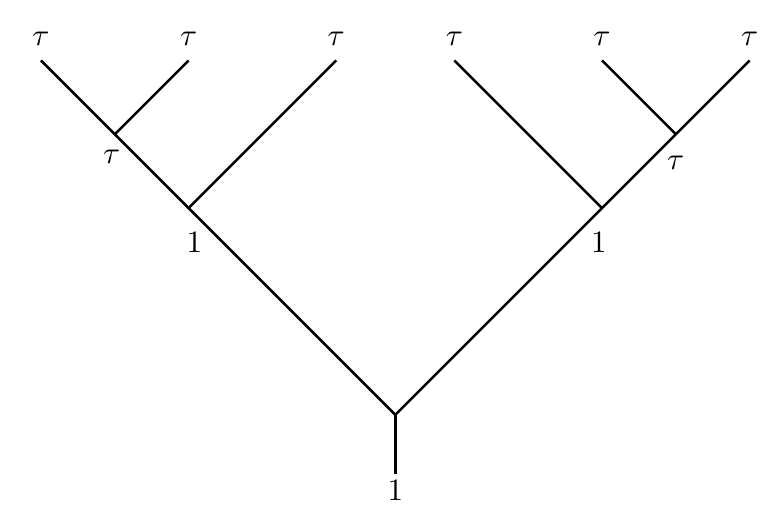}
			\caption{$\ket{NC}$ non-computational state}
		\end{subfigure}
		\caption{Graphical representation of the basis states of two qubits in the six-anyon system.}
		\label{fig_6anyon_basis}
	\end{figure}
	
	Cui et al. \cite{Cui_2Qubit_JoPA2019} presented the encoding of two qubits into the system of six Fibonacci anyons, illustrated in Fig.$~$\ref{fig_6anyon_basis}, and also derived the six-anyon representation of the five braiding operations, $\rho_6(\sigma_k)$. We have verified both the encoding and the braiding and adopted them in this work. The braiding matrices, $\sigma_k$, that braid the $k$ and $k+1$ anyons, in the basis $\{\ket{NC}, \ket{11}, \ket{1\tau}, \ket{\tau 1}, \ket{\tau\tau}\}$ are given by the following:
	\begin{itemize}
		\item $\rho_6(\sigma_1) = R^{\tau\tau}_{\tau} \oplus (R \otimes \ID_2)  $
		\item $\rho_6(\sigma_2) = R^{\tau\tau}_{\tau} \oplus (FRF \otimes \ID_2)  $
		\item $\rho_6(\sigma_3) = P_{14}(R^{\tau\tau}_{\tau} \oplus R \oplus FRF))P_{14}  $
		\item $\rho_6(\sigma_4) = R^{\tau\tau}_{\tau} \oplus (\ID_2 \otimes FRF)  $
		\item $\rho_6(\sigma_5) = R^{\tau\tau}_{\tau} \oplus (\ID_2 \otimes R)  $.
	\end{itemize}
	Here, $R$ and $F$ are the matrices defined in Equations \eqref{eq_Fmat} and \eqref{eq_Rmat} respectively, $P_{14}$ is the permutation matrix obtained by exchanging the first and fourth rows of a $5\times5$ matrix, and $\ID_2$ is the $2\times2$ identity matrix. 
	
	These braiding matrices constitute the \textit{basic gate alphabet} for the topological quantum compilation. In an abuse of notation, we will denote the basic alphabet by $\{\sigma_{1}, \sigma_{2}, \sigma_{3}, \sigma_{4}, \sigma_{5}\}$.
	In addition, we also consider an {\it extended gate alphabet} which includes the inverse braiding matrices and thus consists of ten operations in total, $\{\sigma_{1}, \sigma_{2}, \sigma_{3}, \sigma_{4}, \sigma_{5}, \sigma_{1}^{-1}, \sigma_{2}^{-1}, \sigma_{3}^{-1}, \sigma_{4}^{-1}, \sigma_{5}^{-1} \}$. As is discussed in the following, the extended alphabet impacts the complexity of the topological compilation but does not appear to lead to a significant advantage.
	
	A unitary operation $U$, which results from a product of elements of one of these alphabets, is considered leakage-free if it preserves the 4-dimensional computational subspace $V_C$ \cite{Cui_2Qubit_JoPA2019}. This requires that the $5 \times 5$ matrix, which results from a sequence of elementary braids, can be written as $M \oplus U$ where the matrix $M$ is one-dimensional with the element $M_{11} \in U(1)$ and has unit norm. Since our objective is to approximate a leakage-free gate, we require that the norm of $M_{11}$, given by its absolute value, satisfies $\abs{1 - \abs{M_{11}}} < \delta$ for $\delta \ll 1$. All the gates that constitute both the basic and extended alphabets are leakage-free gates.
	
	
	\section{Approximating the CNOT gate} \label{sec_exactCNOT}
	
	Our first task is to realize the controlled-NOT gate ($\operatorname{CNOT}$) as a product of the elements of one of these alphabets. The $\operatorname{CNOT}$ gate acts on a two-qubit Hilbert space in such a manner that it applies the bit-flip operation, given by the Pauli matrix $\sigma_x$, on the target qubit if the control qubit is in the state $\ket{1}$. We choose the first qubit as the control qubit so that the $\operatorname{CNOT}$ gate can be written in the two-qubit standard computational basis $\{ \ket{00}, \ket{01}, \ket{10}, \ket{11} \}$ as the $4 \times 4$ matrix
	\begin{equation}
		\operatorname{CNOT} = \ID_{2} \oplus \sigma_x =
		\begin{pmatrix}
			1 & 0 & 0 & 0 \\
			0 & 1 & 0 & 0 \\
			0 & 0 & 0 & 1 \\
			0 & 0 & 1 & 0
		\end{pmatrix}.
	\end{equation}
	
	As our system of six Fibonacci anyons corresponds to a five-dimensional Hilbert space, we will be looking for implementations of the $\operatorname{CNOT}$ gate in the form of $M_{11} \oplus A$, with $A$ acting on the four-dimensional computational subspace $V_C$ and satisfies two conditions: (i) it is approximately unitary and (ii) it is `close' to the CNOT gate. 
	
	The first condition requires finding the 17-dimensional subspace of the manifold $U(5)$ which has the total dimension $25$. Specifically, we will first look for the implementations where $\abs{M_{11}} \approx 1$ which defines our first measure of unitarity according to \cite{Cui_2Qubit_JoPA2019}. In general, a more sophisticated measure of unitarity of the matrix $A$ may be desired, for example, in the case that a required implementation has the form $M \oplus A$ where $M$ is more than one-dimensional. In our context, we do not consider the measure given by 
	$\tr(A^\dagger A) /\mbox{dim}(A) = 1$ sufficient, as the tracing operation essentially excludes much information and permits misclassification of unitary/non-unitary matrices.
	
	
	We instead choose to utilize a notion of distance between matrices to construct a measure of unitarity -- the Schatten $p$-norm of the difference between two matrices. The Schatten $p$-norm is given by
	\begin{equation}
		\| A \|_p  \ = \ \tr(|A|^{p})^{1/p}
	\end{equation}
	for a Hermitian matrix $A$, with $1\leq p \leq \infty$, and $\abs{A} = \sqrt{A^\dagger A}$ denotes the operator square root.
	In our case, it suffices to use the trace norm, which corresponds to the $p = 1$ norm, to define a unitarity measure via
	\begin{equation}
		d^U(A) = \| A^\dagger A - \ID\|_1 = \tr \abs{A^\dagger A - \ID},
		\label{eq_unitarymeasure}
	\end{equation}
	In our computations, we calculate both measures of unitarity as they offer a complementary perspective.
	
	For the second condition, we desire the $4\times4$ sub-matrix, constructed in the 4-dimensional computational space $V_C$, to be as close to the $\operatorname{CNOT}$ matrix as possible. To quantify this distance, we shall use the (normalized) Hilbert-Schmidt distance (i.e., the Schatten $p=2$ distance)
	\begin{equation}
		d_2(A,B) = \left\| \frac{A}{\|A\|_2} - \frac{B}{\|B\|_2}\right\|_2,
	\end{equation}
	where $\|.\|_2$ is the Hilbert-Schmidt ($p=2$) norm, and each matrix is normalized by its norm to bound the distance from above. Using this notion of distance, we define the distance of the matrix $A$ to the $\operatorname{CNOT}$ matrix as
	\begin{equation}
		d_2(A) = \left\| \frac{A}{\|A\|_2} - \frac{\operatorname{CNOT}}{\|\operatorname{CNOT}\|_2}\right\|_2 .
		\label{eq_dist_CNOT}
	\end{equation}
	
	Now, the procedure for finding a $4\times4$ sub-matrix $A$ that is both approximately unitary and as close to the $\operatorname{CNOT}$ matrix as possible is as follows:
	\begin{enumerate}
		\item Pick a braid combination length $L$. This will be the number of matrices to take the product of.
		\item For each matrix in the combination, pick one of the 5 braiding matrices $\sigma_k$ of the basic gate alphabet (resp. one of the 10 braiding matrices of the extended alphabet) to use. For a combination of length $L$, there will be a total of $5^L$ (resp. $10^{L}$) combinations to choose from.
		\item Compute the product of the chosen matrices in this combination to obtain the resulting $5\times5$ matrix $M$.
		\item Compute the norm of the $(1,1)$ element ($|M_{11}|$) which serves as the leakage indicator.
		\item Compute the unitary measure $d^U(A)$ of the $4\times4$ sub-matrix $A$.
		\item If the unitarity of the  $4\times4$ sub-matrix $A$ is approximately satisfied within a given threshold, compute the distance $d_2(A)$ between $A$ and the CNOT matrix. Otherwise, repeat the steps 1-5.
		\item Store the combination (`Operator'), distance (`$d_2(A)$'), norm of the $(1,1)$ element (`$|M_{11}|$'), and the unitary measure (`$d^U(A)$').
	\end{enumerate}
	
		
		
		
		
		
		

	\pgfplotstableset{
		alias/No./.initial=Index,
		alias/$L$/.initial=Length,
		alias/$d_2(A)$/.initial=Distance,
		alias/$|M_{11}|$/.initial=Norm11,
		alias/$d^{U}(A)$/.initial=UnitaryMeasure,
	}
	
	\begin{table}[htb!]
		\centering 	
		\begin{filecontents*}{./Data/BestBraids_Unitary_ExactCNOT_maxBr_len31.csv}
Index,Length,Operator,Distance,Norm11,UnitaryMeasure
1,3,111,1.17557,1.0,0.0
2,4,1001,1.0,1.0,0.0
3,5,00000,1.0,1.0,0.0
4,6,443344,1.07589,1.0,0.0
5,7,4334300,0.8972,1.0,0.0
6,8,33333300,0.8957,1.0,0.0
7,9,100100000,1.0,1.0,0.0
8,10,0000000000,1.0,1.0,0.0
9,11,31001000000,0.8972,1.0,0.0
10,12,322022222222,0.8092,1.0,0.0
11,12,011311331113,0.8407,1.0,0.0
12,13,4420410200044,0.8270549054096941,1.0000000000000004,5.246416704584731e-15
13,16,3244411044114244,0.80922,1.0,0.0
14,17,03043034440033034,0.79848,1.0,0.0
15,21,223400344020222234322,0.63212,0.98421,0.05881
16,21,423333044023300323423,0.65249,0.99556,0.0149
17,23,34344031313404411322240,0.78513,0.95385,0.09017
18,27,340113211204400220142121340,0.48414,0.98421,0.09142
19,27,423344224432003320410410133,0.54221,0.98421,0.0626
20,31,2430341014411443041020323332434,0.48225,0.98067,0.07314
21,31,4040232033011013103334443303002,0.69937,0.9899,0.02764
\end{filecontents*}
\pgfplotstabletypeset[
		columns={No.,$L$,Operator,$d_2(A)$,$|M_{11}|$,$d^{U}(A)$}, 
		col sep=comma, 
		column type=R{1cm}, 
		precision=3, 
		every row no 11/.style={after row=\hline}, 
		every head row/.style={before row=\hline\hline,after row=\hline}, 
		every last row/.style={after row=\hline\hline}, 
		columns/No./.style={ 
			string type, column type=L{0.8cm}}, 
		columns/$L$/.style={ 
			string type, column type=L{0.8cm}}, 
		columns/$d^{U}(A)$/.style={column type=R{2.0cm}}, 
		columns/Operator/.style={ 
			string type, column type=l} 
		]{./Data/BestBraids_Unitary_ExactCNOT_maxBr_len31.csv} 
		\caption{Strings of operators from the basic gate alphabet and their corresponding distance $d_2(A)$ to the $\operatorname{CNOT}$ operation, and the norm of the (1,1) element $\abs{M_{11}}$ and $d^U(A)$ which indicate the leakage into the non-computational subspace. The horizontal dividing line separates the results obtained by an exhaustive search and sampling.}
		\label{tab_BestBraids_exact_unitary}
	\end{table}
	
	
	In Table \ref{tab_BestBraids_exact_unitary}, we show the results of searching for an exact CNOT gate using braid sequences of lengths from 3 up to 31 from the basic gate alphabet. 
	Note that in the table, a string "0104230..." indicates that the matrix is made up of the product $\sigma_1\sigma_2\sigma_1\sigma_5\sigma_3\sigma_4\sigma_1..."$.
	The results represent the best outcomes from the total number of trials on the order of $10^{12}$.
	For braid sequences up to length 13, we perform an exhaustive search of all combinations. For braids of length 14 and above, an exhaustive search is intractable and we instead randomly sample the search space in a brute-force manner. 
	The results in Table \ref{tab_BestBraids_exact_unitary}, are the braids that have the lowest distance to the exact $\operatorname{CNOT}$ gate while having a unitary measure less than $0.1$. 
	
	Table \ref{tab_BestBraids_exact_unitary_inv} shows the results of the same compilation problem with the extended gate alphabet which also includes the inverse operations. These operators are labeled as
	\begin{align*}
		\left\{\sigma_{1}, \sigma_{2}, \sigma_{3}, \sigma_{4}, \sigma_{5}, \sigma_{1}^{-1}, \sigma_{2}^{-1}, \sigma_{3}^{-1}, \sigma_{4}^{-1}, \sigma_{5}^{-1} \right\} 
		\to \left\{ 0,\ 1,\ 2,\ 3,\ 4,\ 5,\ 6,\ 7,\ 8,\ 9 \right\}.
	\end{align*}
	The exhaustive search in this case is possible with standard computing resources only for sequences up to the length of 8, beyond which it becomes intractable. We noted that including also the inverse braid operations did not seem to lead to any significant improvement of the convergence to the target $\operatorname{CNOT}$, at least using the available resources.
	
	\pgfplotstableset{
		alias/No./.initial=Index,
		alias/$L$/.initial=Length,
		alias/$d_2(A)$/.initial=Distance,
		alias/$|M_{11}|$/.initial=Norm11,
		alias/$d^{U}(A)$/.initial=UnitaryMeasure,
	}
	
	\begin{table}[htb!]
		\centering 	
		\begin{filecontents*}{./Data/BestBraids_Unitary_Inv_ExactCNOT_maxBr_len24.csv}
Index,Length,Operator,Distance,Norm11,UnitaryMeasure
1,2,30,0.80921668,1.0,0.0
2,3,880,1.16674387,1.0,0.0
3,4,5300,0.80921668,1.0,0.0
4,5,48440,0.97082132,1.0,0.0
5,6,553000,0.80921668,1.0,0.0
6,7,4334300,0.89720298,1.0,0.0
7,8,37770000,0.79160802,0.95385012,0.09016994
8,9,441001048,0.81908,1.0,0.0
9,10,2555337582,0.79161,0.95385,0.09017
10,11,59672955693,0.81908,1.0,0.0
11,12,374050007970,0.79161,0.95385,0.09017
12,12,306947382105,0.80922,1.0,0.0
13,13,6595888969003,0.79848,1.0,0.0
14,14,75139699757375,0.64279,0.98421,0.05881
15,20,78450807782427077787,0.6145,0.97591,0.09385
16,24,287430798424700981936977,0.54975,0.98421,0.05881
\end{filecontents*}
\pgfplotstabletypeset[
		columns={No.,$L$,Operator,$d_2(A)$,$|M_{11}|$,$d^{U}(A)$}, 
		col sep=comma, 
		column type=R{1.1cm}, 
		precision=4, 
		every row no 6/.style={after row=\hline}, 
		every head row/.style={before row=\hline\hline,after row=\hline}, 
		every last row/.style={after row=\hline\hline}, 
		columns/No./.style={ 
			string type, column type=L{1cm}}, 
		columns/$L$/.style={ 
			string type, column type=L{1cm}}, 
		columns/Operator/.style={ 
			string type, column type=l}, 
		columns/$d^{U}(A)$/.style={column type=R{1.5cm}},
		]{./Data/BestBraids_Unitary_Inv_ExactCNOT_maxBr_len24.csv} 
		\caption{Strings of operators from the extended gate alphabet and their corresponding distance $d_2(A)$ to the CNOT operation, and the norm of the (1,1) element $\abs{M_{11}}$ and $d^U(A)$ which indicate the leakage into the non-computational subspace. The horizontal dividing line separates the results obtained by exhaustive search and sampling.}
		\label{tab_BestBraids_exact_unitary_inv}
	\end{table}
	
	We see from the results in Tables \ref{tab_BestBraids_exact_unitary} and \ref{tab_BestBraids_exact_unitary_inv}, that the search for an approximate implementation of the $\operatorname{CNOT}$ operation fails to reach the optimization target within a reasonable error $\epsilon \ll 1$ while satisfying the unitarity condition. It seems plausible that the CNOT operation is not a natural target for sequences taken from either the basic or extended gate alphabet. Clearly, longer sequences of elementary braids need to be employed in the implementation of the $\operatorname{CNOT}$ gate in the Fibonacci anyon system considered here. In other words, finding, or at least approaching a small neighborhood of a single point, such as the $\operatorname{CNOT}$ gate, in the 16-dimensional manifold of $U(4)$ using five or ten discrete operations, most of them from the subgroup $U(2) \otimes U(2)$ of single-qubit operations, may not have an easy solution. In the following, we will focus on a different approach to implement two-qubit operations such as the $\operatorname{CNOT}$ gate with the Fibonacci anyons which is considerably more efficient.
	
	
	\section{Approximating the local equivalence class [CNOT]}\label{sec_class}
	
	Now, instead of looking for a matrix that is as close as possible to the $\operatorname{CNOT}$ gate, we will instead search for any two-qubit operation that is approximately equivalent to the $\operatorname{CNOT}$ gate up to any single qubit transformations. The set of all operations equivalent in this sense to the CNOT gate forms the local equivalence class $[\operatorname{CNOT}]$. The local equivalence class is uniquely characterized by three real parameters called local invariants, first introduced by Makhlin \cite{Makhlin_Invariants_QIP2002}. These were used, in combination with the Cartan decomposition of the Lie group $SU(4)$, by Zhang et al. \cite{ZhangValaBirgitta_GeometricTheory_PRA2003} for the development of the geometric theory of two-qubit operations which forms the basis for our considerations here. 
	
\begin{figure}[htb!]
		\centering
		\begin{subfigure}{0.35\linewidth}
			\centering
			\includegraphics[width=1\linewidth]{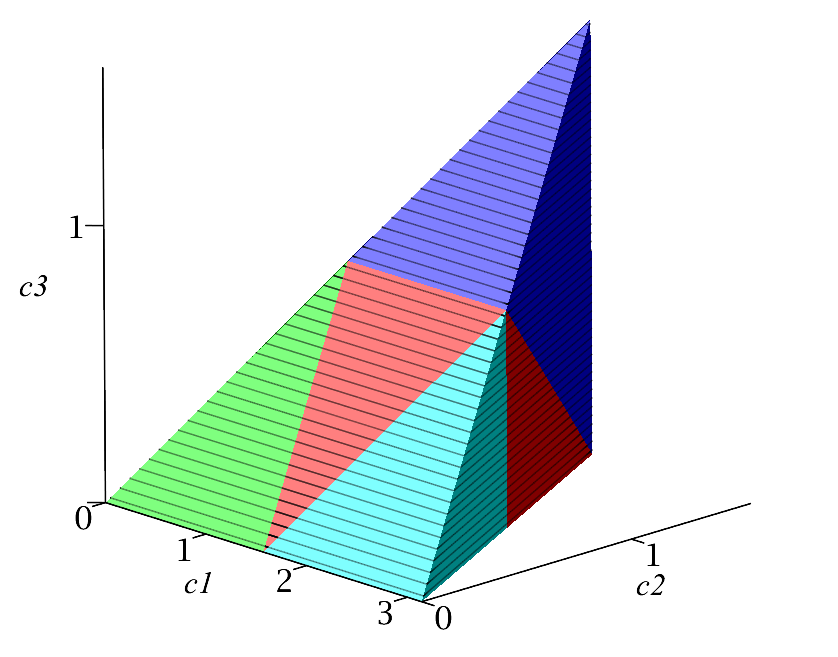}
			\caption{}
		\end{subfigure}
		\hspace*{0.5cm}
		\begin{subfigure}{0.35\linewidth}
			\centering
			\includegraphics[width=1\linewidth]{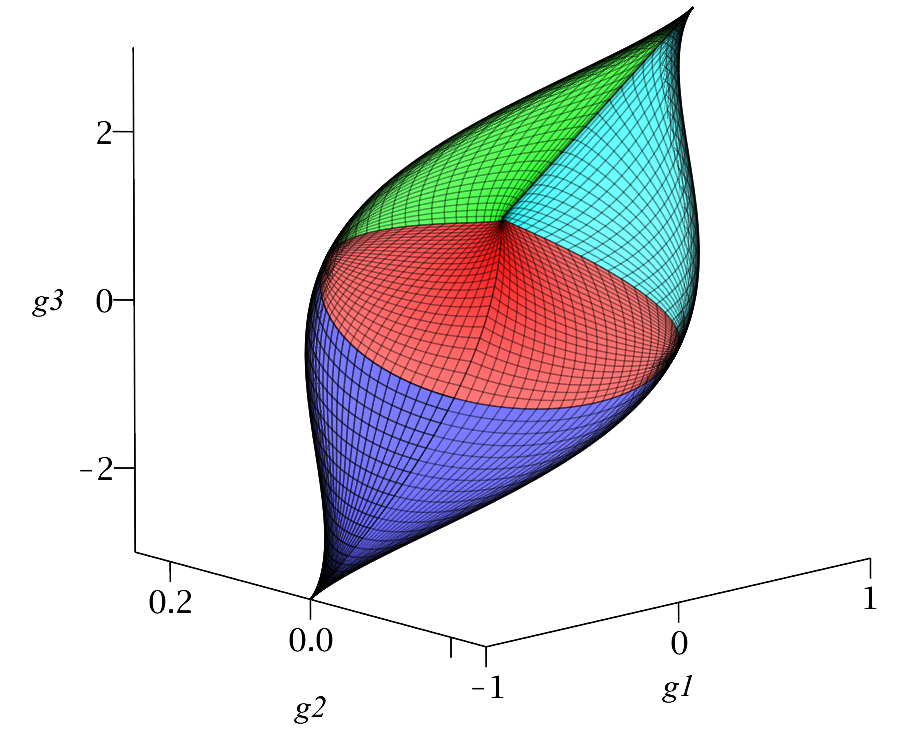}
			\caption{}
		\end{subfigure}
		\caption{Illustration of the Weyl chamber \cite{Watts_15} (a) in the representation given by the coefficients of the elements of the Cartan subalgebra $c_1,c_2,c_3$, and (b) in the representation given by the local invariants $g_1,g_2,g_3$.}
		\label{fig_weyl_chamber}
	\end{figure}

	Employing the local equivalence class of a two-qubit gate instead of the gate itself significantly expands the optimization target \cite{Muller_11} which now corresponds to a point in a three-dimensional space of local equivalence classes called the Weyl chamber. This space is defined either by the coefficients $c_1, c_2, c_3$ of a corresponding element of the Cartan subalgebra, with certain Weyl reflection symmetry removed, or by the local invariants $g_1, g_2, g_3$. Both representations of the set of local equivalence classes are illustrated in Fig. \ref{fig_weyl_chamber}. The procedure for finding the desired local equivalence class is essentially the same as in Section \ref{sec_exactCNOT}. We take a product of the elementary braiding matrices $\sigma_k$  and first test for the unitarity of the overall operation in the 4-dimensional subspace. If this condition is satisfied, we search for an element of the local equivalence class $[\operatorname{CNOT}]$ using a cost function defined in terms of the local invariants.
	
	\subsection{Cost function}
	We briefly review the construction of the local invariants and then define the relevant cost function. A two-qubit unitary operation $U \in SU(4)$ has the Cartan decomposition given by $U = k_1 \mathcal{A} k_2$ where $k_1, k_2 \in SU(2) \otimes SU(2)$ and $\mathcal{A}$ is generated by an element of the Cartan subalgebra of the algebra $su(4)$ spanned by the generators proportional to the tensor products of the Pauli matrices $\sigma_x \otimes \sigma_x,\ \sigma_y \otimes \sigma_y,\ \sigma_z \otimes \sigma_z$.  
	We first transform $U$ from the standard computational basis to the Bell basis,
	\begin{equation}
		\mathcal{Q} = \frac{1}{\sqrt{2}}
		\begin{pmatrix}
			1 & 0 & 0 & i \\
			0 & i & 1 & 0 \\
			0 & i & -1 & 0 \\
			1 & 0 & 0 & -i
		\end{pmatrix}, 
		\quad \implies \quad U_B = \mathcal{Q}^{\dagger} U \mathcal{Q}.
	\end{equation}
	In this representation, the local factors $k_1$ and $k_2$ of the Cartan decomposition become respectively elements $O_1$ and $O_2$ of the orthogonal group with the property $O^TO = OO^T = \ID$, and the non-local factor  $\mathcal{Q}^\dagger \mathcal{A}\mathcal{Q}$  becomes diagonal. To define the Makhlin matrix, we take the following product
	\begin{equation}
		m_U = U_B^{T}~ U_B \ ,
	\end{equation}
	which eliminates one of the local operations. The other set of local operations in fact diagonalizes the Makhlin matrix $m_U$ to produce the diagonal $\mathcal{Q}^\dagger A \mathcal{Q}$. From the relevant characteristic equation, one can derive the local invariants in the following form:
	\begin{align}
		g_1 = \frac{\text{Re}\{\tr^2(m_U)\}}{16} , \quad
		g_2 = \frac{\text{Im}\{\tr^2(m_U)\}}{16} , \quad
		g_3 = \frac{\tr^2(m_U) - \tr(m_U^2)}{4}. \label{eq_gs}
	\end{align}
	The values of these invariants for various equivalence classes are given in Table \ref{tab_gs}.
	
	\begin{table}[htb!]
		\centering
		\begin{tabular}{l | r r r}
			\hline
			\hline
			Class \quad \quad & \quad $g_1$ \quad & \quad $g_2$ \quad & \quad $g_3$ \quad \\ 
			\hline
			$[\ID]$ \quad \quad & \quad $1$ \quad & \quad $0$ \quad & \quad $3$ \quad \\
			$[\operatorname{DCNOT}]$ \quad \quad & \quad $0$ \quad & \quad $0$ \quad & \quad $-1$ \quad \\
			$[\operatorname{SWAP}]$ \quad \quad & \quad $-1$ \quad & \quad $0$ \quad & \quad $-3$ \quad \\
			$[\operatorname{B}]$ \quad \quad & \quad $0$ \quad & \quad $0$ \quad & \quad $0$ \quad \\
			$[\operatorname{CNOT}]$ \quad \quad & \quad $0$ \quad & \quad $0$ \quad & \quad $1$ \quad \\
			$[\sqrt{\operatorname{SWAP}}]$ \quad & \quad $0$ \quad & \quad $1/4$ \quad & \quad $0$ \quad \\
			\hline
			\hline
		\end{tabular}
		\caption{Values of the Makhlin invariants for different local equivalence classes.}
		\label{tab_gs}
	\end{table}
	
	These invariants are defined for the elements of the Lie group $SU(4)$. In order to generalise these to elements of $U(4)$, we express $U = e^{i\alpha} \tilde{U} \in U(4)$, where $\tilde{U} \in SU(4)$. The global phase, $e^{i\alpha}$ can be eliminated by dividing the local invariants Eq.~\eqref{eq_gs} by $\det(U) = e^{4i\alpha}$. Thus, the final form of the invariants, for elements of $U(4)$, are
	\begin{align}
		g_1 = \text{Re}\left\{\frac{\tr^2(m_U)}{16\cdot \det(U)}\right\} , \quad
		g_2 = \text{Im}\left\{\frac{\tr^2(m_U)}{16\cdot \det(U)}\right\}, \quad
		g_3 = \frac{\tr^2(m_U) - \tr(m_U^2)}{4\cdot \det(U)}. \label{eq_gs_det}
	\end{align}
	Note that since $U \in U(4)$, $g_1,g_2,g_3 \in \mathbb{R}$ as stated earlier.
	
	While the full $5\times5$ matrices that we obtain from an arbitrary sequence of $\sigma_k$ will be unitary, the $4\times4$ sub-matrix will generally be only approximately unitary. As a result, $g_3$ can have complex values. As previously, we first test the unitarity of the resulting $4\times4$ sub-matrix and only if it is satisfied to the required accuracy, do we proceed to the calculation of the local invariants. In these cases, the imaginary component of $g_3$ does not exceed the numerical value of $10^{-5}$.
	
	We now define a distance measure to a given local equivalence class using the local invariants, namely, we define the distance of a matrix $U$ to an equivalence class $\mathbf{E}$ as
	\begin{equation}
		d^{\mathbf{E}}(U) \ = \ \sum_{i=1}^{3}\Delta g_i^2 \quad , \quad \Delta g_i \ = \ |g_i(\mathbf{E}) - g_i(U)|.
	\end{equation}
	We are now prepared to search for sequences of braiding matrices, intending to approximate the $\operatorname{CNOT}$ local equivalence class. In the following discussion, we denote the distance of an approximately unitary $4\times 4$ matrix $A$ from the local equivalence class $[\operatorname{CNOT}]$ as $d(A) = d^{\mathbf{CNOT}}(A)$.
	
	
	\pgfplotstableset{
		alias/No./.initial=Index,
		alias/$L$/.initial=Length,
		alias/$d(A)$/.initial=Distance,
		alias/$|M_{11}|$/.initial=Norm11,
		alias/$d^{U}(A)$/.initial=UnitaryMeasure,
	}
	
	\begin{table}[htb!]
		\centering 	
		\begin{filecontents*}{./Data/TrialBraidsDF_Unitary_CNOTEquiv_TopforMaxBrL39.csv}
Index,Length,Operator,Distance,Norm11,UnitaryMeasure
1,3,000,5.0,1.0,0.0
2,4,0000,5.0,1.0,0.0
3,5,00000,5.0,1.0,0.0
4,6,222000,2.9082,0.95385,0.09017
5,7,0000000,5.0,1.0,0.0
6,8,00000000,5.0,1.0,0.0
7,9,000000000,5.0,1.0,0.0
8,10,2221001222,0.46345,0.97591,0.0476
9,11,22210012220,0.46345,0.97591,0.0476
10,12,223443100122,2.0e-5,0.97591,0.0476
11,13,2213404301242,2.0e-5,0.97591,0.0476
12,17,34224334310031224,2.0e-5,0.97591,0.0476
13,21,242314040312211412221,0.0943398,0.991554,0.0292798
14,24,422211222020214000112021,0.007829,0.994801,0.0177622
15,27,101422102223130431111322222,0.0012317,0.994225,0.0220096
16,29,31423322434332420442310301422,0.000106,0.995169,0.0184123
17,30,222223043422422024043333320003,7.779e-9,0.998025,0.0075642
18,31,2344130440331032213334400344312,2.2e-8,0.997344,0.0094756
19,33,423330314001220244224333334034032,7.8e-9,0.9980252,0.0068292
20,39,132424244140142042111112011202122020400,0.0002057958,0.9994728,0.0020892
21,39,231333001244422220433403402422343302043,7.8e-9,0.9980252,0.0047446
\end{filecontents*}
\pgfplotstabletypeset[
		columns={No.,$L$,Operator,$d(A)$,$|M_{11}|$,$d^{U}(A)$}, 
		col sep=comma, 
		column type=R{1.1cm}, 
		precision=3, 
		every row no 10/.style={after row=\hline}, 
		every head row/.style={before row=\hline\hline,after row=\hline}, 
		every last row/.style={after row=\hline\hline}, 
		columns/No./.style={ 
			string type, column type=L{1.1cm}}, 
		columns/$L$/.style={ 
			string type, column type=L{1.1cm}}, 
		columns/Operator/.style={ 
			string type, column type=l}, 
		columns/$d(A)$/.style={column type=R{2cm}},
		columns/$d^{U}(A)$/.style={column type=R{2cm}},
		]{./Data/TrialBraidsDF_Unitary_CNOTEquiv_TopforMaxBrL39.csv},
		\caption{String of operators and their corresponding distance $d$ to the local equivalence class $[\operatorname{CNOT}]$, the norm of the (1,1) element $\abs{M_{11}}$, and the unitary measure $d^U$. The horizontal dividing line separates the results obtained by an exhaustive search and sampling. 
		}
		\label{tab_braids_CNOTequiv_unitary}
	\end{table}
	
	\subsection{Results}
	
	We first look for sequences of $\sigma_k$, whose $4\times 4$ sub-matrix is approximately unitary over the computational subspace. Then we compute the distance $d(A)$ of the matrix $A$ to the local equivalence class $[\operatorname{CNOT}]$.
	Table \ref{tab_braids_CNOTequiv_unitary} shows the results for the basic gate alphabet. We perform an exhaustive search up to length 14, and randomly sample for longer sequences.
	It is evident from the results that there is a trade-off between having a unitary matrix and having a matrix that is close to the $\operatorname{CNOT}$ local equivalence class (in terms of the invariants). Nevertheless, one can achieve very small distances while maintaining a very good approximation to unitarity by increasing the length of the chain. Table \ref{tab_BestBraids_equiv_unitary_inv} shows the compilation results obtained for the extended gate alphabet which includes also inverse operations. In this case, we have only been able to perform the exhaustive search for sequences of length up to 8. We observe no obvious difference between the results obtained for the basic or extended alphabet, though we acknowledge that a more comprehensive search may need to be employed for such a comparison. This is however beyond our present objectives which reside in comparison between topological compilation with respect to a specific entangling operation and compilation with respect to its local equivalence class. It is evident that the latter results in significant improvements in terms of convergence as well as the length or number of elementary operations required for the implementation of an entangling operation from the local equivalence class $[\operatorname{CNOT}]$. It is of course plausible that other local equivalence classes may be generated from our gate alphabet more naturally, as we shall discuss next.
	
	\pgfplotstableset{
		alias/No./.initial=Index,
		alias/$L$/.initial=Length,
		alias/$d(A)$/.initial=Distance,
		alias/$|M_{11}|$/.initial=Norm11,
		alias/$d^{U}(A)$/.initial=UnitaryMeasure,
	}
	
	\begin{table}[htb!]
		\centering 	
		\begin{filecontents*}{./Data/TrialBraidsDF_Unitary_CNOTEquiv_inv_TopforMaxBrL21.csv}
Index,Length,Operator,Distance,Norm11,UnitaryMeasure
1,1,0,5.0,1.0,0.0
2,2,00,5.0,1.0,0.0
3,3,222,2.9082,0.95385,0.09017
4,4,2220,2.9082,0.95385,0.09017
5,5,22200,2.9082,0.95385,0.09017
6,6,222000,2.9082,0.95385,0.09017
7,7,2220000,2.9082,0.95385,0.09017
8,10,6577100177,0.46345,0.97591,0.0476
9,12,208923329892,2.0e-5,0.97591,0.0476
10,13,3734788743705,2.0e-5,0.97591,0.0476
11,14,26952811352642,2.0e-5,0.97591,0.0476
12,17,76177322244377075,0.06785,0.98383,0.04974
13,19,9777577842888487783,0.01579,0.99435,0.01784
14,20,13282844544028208235,0.00116,0.98401,0.04577
15,21,726202360623786904622,0.00077,0.98753,0.048
16,21,524546245839171526261,2.0e-5,0.97591,0.0476
\end{filecontents*}
\pgfplotstabletypeset[
		columns={No.,$L$,Operator,$d(A)$,$|M_{11}|$,$d^{U}(A)$}, 
		col sep=comma, 
		column type=R{1.2cm}, 
		precision=3, 
		every row no 6/.style={after row=\hline}, 
		every head row/.style={before row=\hline\hline,after row=\hline}, 
		every last row/.style={after row=\hline\hline}, 
		columns/Operator/.style={ 
			string type, column type=l}, 
		columns/No./.style={ 
			string type, column type=L{1.2cm}}, 
		columns/$L$/.style={ 
			string type, column type=L{1.2cm}}, 
		columns/$d(A)$/.style={column type=R{2.2cm}},
		columns/$d^{U}(A)$/.style={column type=R{2.2cm}},
		]{./Data/TrialBraidsDF_Unitary_CNOTEquiv_inv_TopforMaxBrL21.csv} 
		\caption{String of operators from the extended gate alphabet and their corresponding distance to the CNOT equivalence class ($d$), and the norm of the $(1,1)$ element, and the unitary measure $d^U$. The horizontal dividing line separates the results obtained by an exhaustive search and sampling. 
		}
		\label{tab_BestBraids_equiv_unitary_inv}
	\end{table}

	
	\section{Natural local equivalence classes}\label{sec_natural}
	
	We now aim to identify which, if any, local equivalence classes can be naturally generated by braiding Fibonacci anyons. Since we do not observe a significant difference between the results obtained with the basic and extended alphabet, we use the former which allows us to explore longer sequences via an exhaustive search approach, and also sample larger fractions of the total space in the cases where exhaustive searches are intractable. Since we perform our computations using the local invariants, we plot our results in the Weyl chamber in the representation given by $g_1,g_2,g_3$. As before, our first step is to test the unitarity of the $4 \times 4$ sub-matrix. If this is satisfied, we evaluate the local invariants and locate the relevant class in the Weyl chamber. 
	Table \ref{tab_braids_invars} shows the unique invariants for resulting matrices that are close to unitary for combinations of lengths up to 39. For the sequences presented therein, $\operatorname{Im}(g_3)  < 10^{-5}$, and hence it is omitted from the table.

	\pgfplotstableset{
		alias/No./.initial=Index,
		alias/$L$/.initial=Length,
		alias/$g_1$/.initial=g1,
		alias/$g_2$/.initial=g2,
		alias/${Re}(g_3)$/.initial=g3Re,
		alias/${Im}(g_3)$/.initial=g3Im,
		alias/$|M_{11}|$/.initial=Norm11,
		alias/$d^{U}(A)$/.initial=UnitaryMeasure,
	}
	
	\begin{table}[htb!]
		\centering 	
		\begin{filecontents*}{./Data/Braids_UnitaryInvars_MaxBr39.csv}
Index,Length,Operator,Distance,Norm11,UnitaryMeasure,g1,g2,g3Re,g3Im
1,3,000,5.0,1.0,0.0,1.0,-0.0,3.0,-0.0
2,4,0000,5.0,1.0,0.0,1.0,0.0,3.0,0.0
3,5,00000,5.0,1.0,0.0,1.0,-0.0,3.0,-0.0
4,6,222000,2.9082,0.95385,0.09017,0.7623883,0.020126,2.5247766,0.040252
5,7,0000000,5.0,1.0,0.0,1.0,-0.0,3.0,0.0
6,8,00000000,5.0,1.0,0.0,1.0,-0.0,3.0,-0.0
7,9,000000000,5.0,1.0,0.0,1.0,-0.0,3.0,0.0
8,10,2221001222,0.46345,0.97591,0.0476,0.3042447,0.0112204,1.6084894,0.0224407
9,11,22210012220,0.46345,0.97591,0.0476,0.3042447,0.0112204,1.6084894,0.0224407
10,12,223443100122,2.0e-5,0.97591,0.0476,0.0017861,0.0010714,1.0035722,0.0021429
11,13,2213404301242,2.0e-5,0.97591,0.0476,0.0017861,0.0010714,1.0035722,0.0021429
12,17,34224334310031224,2.0e-5,0.97591,0.0476,0.0017861,0.0010714,1.0035722,0.0021429
13,21,242314040312211412221,0.0943398,0.991554,0.0292798,0.1373594,-0.0005856,1.2747189,-0.0011712
14,24,422211222020214000112021,0.007829,0.994801,0.0177622,0.0395622,-0.0007907,1.0791245,-0.0015814
15,27,101422102223130431111322222,0.0012317,0.994225,0.0220096,0.015693,-0.0002556,1.031386,-0.0005112
16,29,31423322434332420442310301422,0.000106,0.995169,0.0184123,0.0045935,0.0003095,1.009187,0.0006191
17,30,222223043422422024043333320003,7.779e-9,0.998025,0.0075642,3.88e-5,-7.3e-6,1.0000775,-1.46e-5
18,31,2344130440331032213334400344312,2.2e-8,0.997344,0.0094756,6.39e-5,1.76e-5,1.0001278,3.52e-5
19,33,423330314001220244224333334034032,7.8e-9,0.9980252,0.0068292,3.88e-5,-7.3e-6,1.0000775,-1.46e-5
20,39,132424244140142042111112011202122020400,0.0002057958,0.9994728,0.0020892,0.0064154,3.64e-5,1.0128309,7.28e-5
21,39,231333001244422220433403402422343302043,7.8e-9,0.9980252,0.0047446,3.88e-5,-7.3e-6,1.0000775,-1.46e-5
\end{filecontents*}
\pgfplotstabletypeset[
		columns={No.,$L$,Operator,$g_1$,$g_2$,${Re}(g_3)$},
	col sep=comma, 
	column type=R{1cm}, 
	precision=2, 
	every row no 10/.style={after row=\hline\vspace{.1cm}}, 
	every head row/.style={before row=\hline\hline,after row=\hline}, 
	every last row/.style={after row=\hline\hline}, 
	columns/No./.style={ 
		string type, column type=L{1.3cm} }, 
	columns/$L$/.style={ 
		string type, column type=L{1.3cm} }, 
	columns/Operator/.style={ 
		string type, column type=l}, 
	columns/$g_1$/.style={column type=R{2.2cm}},
	columns/$g_2$/.style={column type=R{2.2cm}},
	columns/${Re}(g_3)$/.style={column type=R{1.4cm}},
	]{./Data/Braids_UnitaryInvars_MaxBr39.csv}
	\caption{Operators (equivalent to Table \ref{tab_braids_CNOTequiv_unitary}) and their corresponding invariants $g_1,g_2,g_3$. 
		The horizontal dividing line separates the results obtained by an exhaustive search and sampling.
	}
	\label{tab_braids_invars}
\end{table}

\begin{figure}[htb!]
	\centering
	\begin{subfigure}{0.4\linewidth}
		\centering
		\includegraphics[width=1\linewidth]{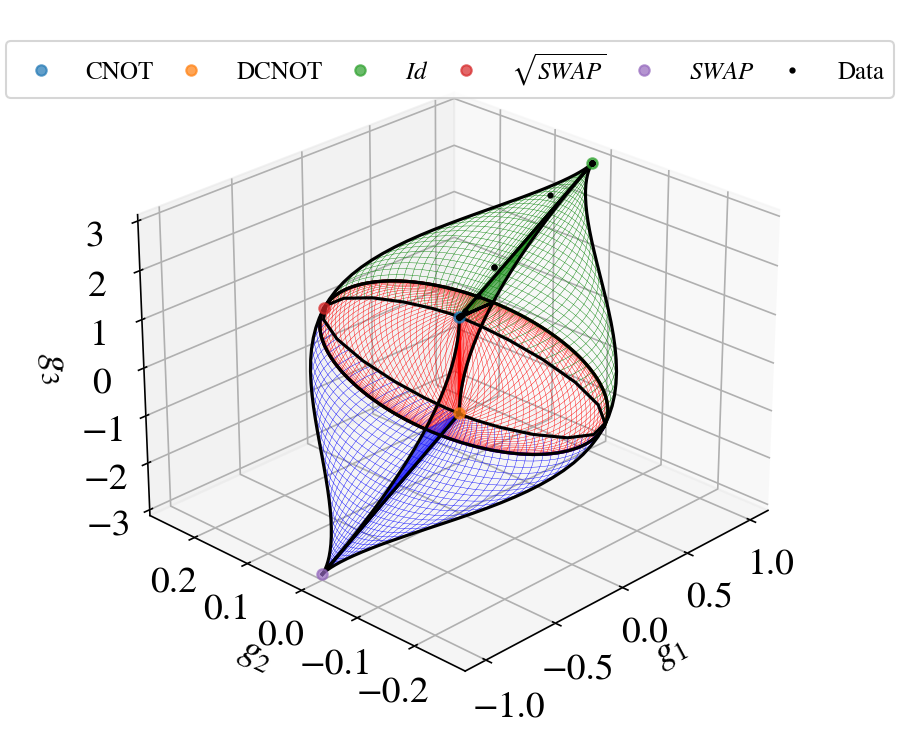}
		\caption{Data plotted alongside the full Weyl chamber.}
	\end{subfigure}
	\begin{subfigure}{0.4\linewidth}
		\centering
		\includegraphics[width=1\linewidth]{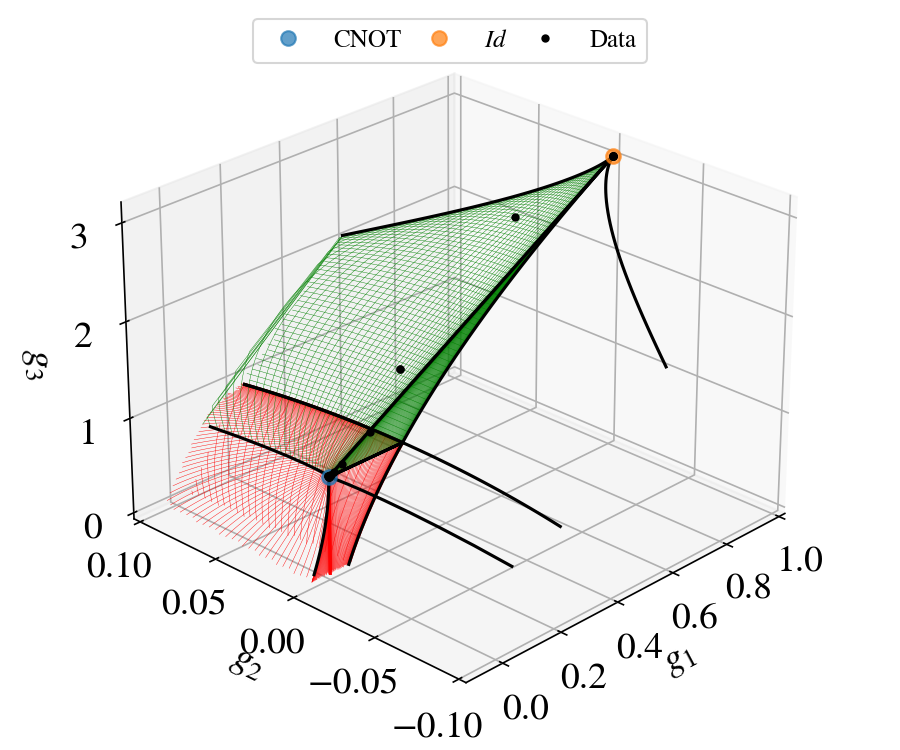}
		\caption{Data plotted alongside the full Weyl chamber, focused on the top half of the Weyl chamber.}
	\end{subfigure}
	\caption{Invariants of matrices from Table \ref{tab_braids_invars} plotted in $g_1,g_2,g_3$ space (braids of length 3-39).}
	\label{fig_invars_CNOT_res}
\end{figure}

Although a comparison could be made between the values in this table and those in Table \ref{tab_gs}, we find that a graphical representation offers a clearer illustration of our results.
In Fig.~\ref{fig_invars_CNOT_res}, we first plot the invariants for the matrices that are shown in Table \ref{tab_braids_invars}, and also illustrate the Weyl chamber in the invariant space, highlighting the perfect entangler region in red. The relevant equivalence classes are highlighted as points on the vertices of the chamber. Since the matrices in the table above a braid of length 12 have a small distance to the CNOT local equivalence class, the data points are generally concentrated around the $[\operatorname{CNOT}]$ point in the Weyl chamber.

\begin{figure}[h!]
	\centering
	\begin{subfigure}{0.45\linewidth}
		\centering
		\includegraphics[width=1\linewidth]{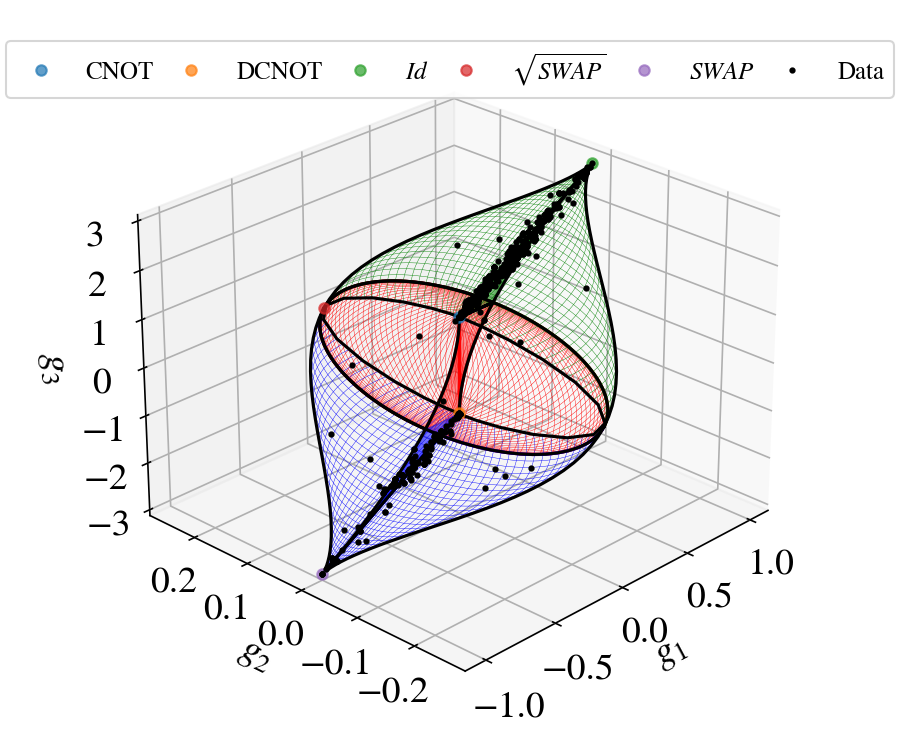}
		\caption{Data plotted alongside the full Weyl chamber.}
	\end{subfigure}
	\begin{subfigure}{0.45\linewidth}
		\centering
		\includegraphics[width=1\linewidth]{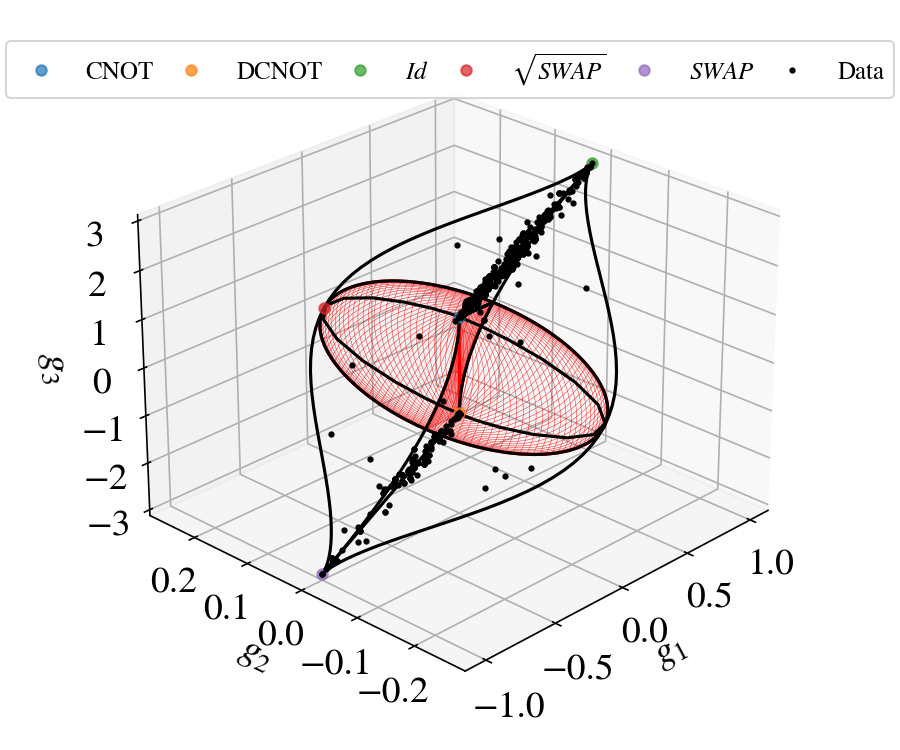}
		\caption{Data plotted alongside the perfect entangler section of the Weyl chamber -- coined the `Eye of Sauron' in \cite{WattsVala_PerfectEntanglers_Ent2013}.}
	\end{subfigure}
	\begin{subfigure}{0.45\linewidth}
		\centering
		\includegraphics[width=1\linewidth]{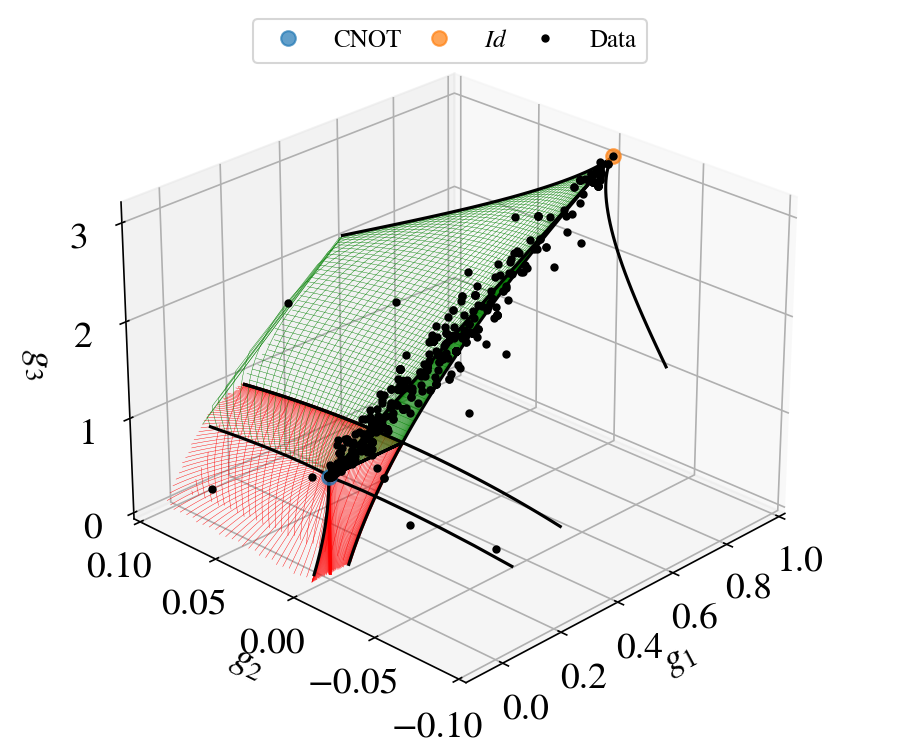}
		\caption{Data plotted alongside the full Weyl chamber, focused on the top half of the Weyl chamber.}
	\end{subfigure}
	\begin{subfigure}{0.45\linewidth}
		\centering
		\includegraphics[width=1\linewidth]{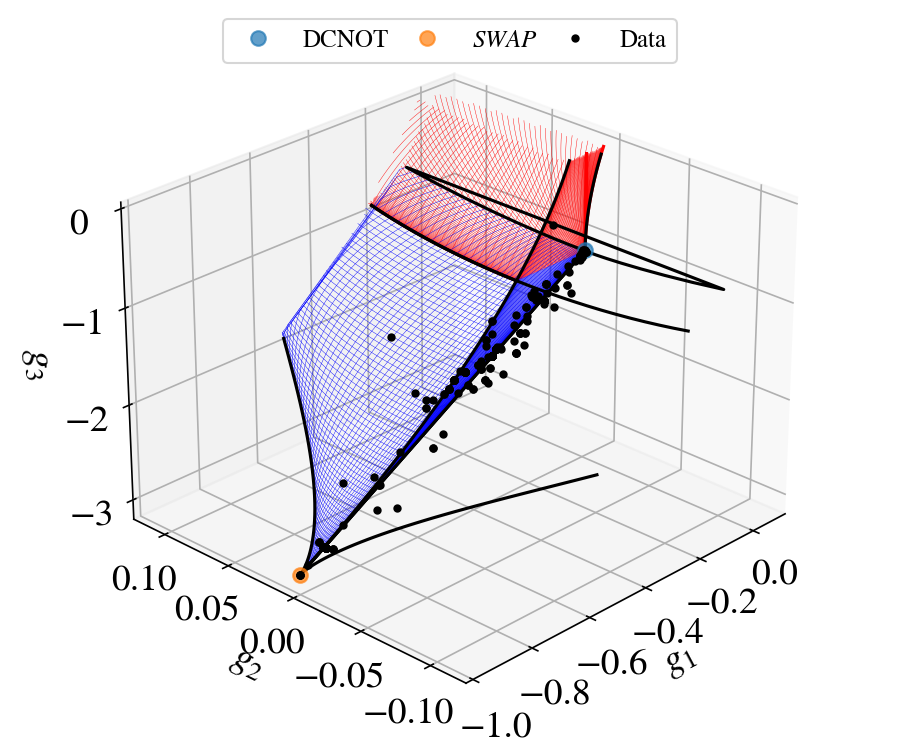}
		\caption{Data plotted alongside one half of the Weyl chamber, focused on the $\operatorname{CNOT}$ point in the space.}
	\end{subfigure}
	\caption{Invariants of random combinations ($\sim 1000$) resulting in approximately unitary matrices plotted in $g_1,g_2,g_3$ space - Braids of length 40.}
	\label{fig_invars_rand_uni}
\end{figure}

Other unitary operations, which we can obtain from sequences of the elementary braids and which are not close to $[\operatorname{CNOT}]$, are particularly interesting as they provide insight into what type of entangling operations are naturally generated in our anyonic system. In Figure \ref{fig_invars_rand_uni} we plot the matrices found that approximately satisfy the unitarity condition following a random sampling of the combinations of length 40. We observe that these natural operations lie between the local equivalence classes $[\ID]$ and $[\operatorname{CNOT}]$, and the classes $[\operatorname{DCNOT}]$ and $[\operatorname{SWAP}]$ for the length of braid sequences investigated here.

Now that we have seen there are other equivalence classes that our combinations of $\sigma_k$ flow to, we shall determine the distance to the closest equivalence class of some of the selected two-qubit operations, such as $\operatorname{CNOT}$, double-$\operatorname{CNOT}$ or $\operatorname{DCNOT}$, $\operatorname{SWAP}$ and the identity operation $\ID$, in terms of the local invariants.
In Table \ref{tab_closest_braids} we show the results for some of the braiding matrices in Figure \ref{fig_invars_rand_uni}. Note that we only show a small sample of the $\sim1000$ points. 
From this table, we observe that we can achieve a perfectly unitary matrix that is locally equivalent to a $\operatorname{SWAP}$ gate. We provide an example of the $5\times 5$ matrix $[M_{11} \oplus A]$ which is locally equivalent to the $\operatorname{SWAP}$ gate in Eq.~\eqref{eq_Mat_SWAP_40}. 
\begin{equation}
	M_{11} \oplus A = 
	\begin{pmatrix}
		-1.0 + 0.0i & +0.00  + 0.00i & +0.00 + 0.00i & +0.00 + 0.00i & +0.00 + 0.00i \\
		+0.0 + 0.0i & -0.04 - 0.31i & -0.36 + 0.18i & +0.49 + 0.22i & +0.07 - 0.67i \\
		+0.0 + 0.0i & +0.49 - 0.22i & -0.46 - 0.50i & +0.04 - 0.31i & -0.39 + 0.07i \\
		+0.0 + 0.0i & -0.39 - 0.07i & -0.04 - 0.31i & +0.46 - 0.50i & +0.49 + 0.22i \\
		+0.0 + 0.0i & -0.07 - 0.67i & -0.49 - 0.22i & -0.36 - 0.18i & +0.04 - 0.31i 
	\end{pmatrix}
	\label{eq_Mat_SWAP_40}
\end{equation}
We see that the element $M_{11}$ is, within the numerical accuracy, equal to $1$. The form of the sub-matrix A is clearly distinct from the form of the $\operatorname{SWAP}$ gate in the standard representation, however, the two are locally equivalent up to single-qubit transformations.

\pgfplotstableset{
	alias/No./.initial=Index,
	alias/$L$/.initial=L,
	alias/$d(A)$/.initial=d,
	alias/$|M_{11}|$/.initial=Norm11,
	alias/$d^{U}(A)$/.initial=Unitarity,
}

\begin{table}[htb!]
	\centering 	
	\begin{filecontents*}{./Data/BraidsDF_Unitary_CloesestEquiv_TopforMaxBrL40.csv}
Index,L,Operator,d,Class,Norm11,Unitarity
1,40,0131130443333410041340121001420341430104,0.0,ID,1.0000000000000013,2.43e-14
2,40,1332041020222120241220324100023023433334,0.0,SWAP,0.9999999999999994,2.0e-14
3,40,1433031222342422203023223230240443302034,4.29409e-11,CNOT,0.9886792534028116,0.0447395645609355
4,40,2042023223234230444023402233310403144302,2.19671883e-8,CNOT,0.9973439565552524,0.0094755907922692
5,40,2111210000021122401344030030343310112200,2.77751663e-8,CNOT,0.9903742263574016,0.0366381652968268
6,40,2000011042011221000220214134423024141113,4.9783036945e-6,CNOT,0.9846674433401754,0.0394426051186044
7,40,4424112204224004142400202443114301430203,1.21973484919e-5,CNOT,0.9916948769184808,0.048256565499393
8,40,4123222404221021243042034403104041340422,2.16908552831e-5,DCNOT,0.9759111274880214,0.0475974712450693
9,40,4111441434443422140131014432210304341300,2.16908552831e-5,CNOT,0.9759111274880216,0.0475974712450814
10,40,3214304124313214030210210410322210013242,9.85846104585e-5,DCNOT,0.9920644466677628,0.0316097723905393
11,40,3010240424120241112044202220001020142122,0.0002985380251758,CNOT,0.9909886785407772,0.0348216248773324
12,40,2303323004100444134324422204022120202124,0.0012316740873647,CNOT,0.9942251896059562,0.0346177959827376
13,40,4403124432032044210003142322243041131112,0.0012316740873647,CNOT,0.9942251896059566,0.0433971241203069
14,40,1442024304110331441224310343121400044432,0.0012316740873648,DCNOT,0.9942251896059562,0.0433971241203085
\end{filecontents*}
\pgfplotstabletypeset[
	columns={No.,Operator,$d(A)$,Class,$|M_{11}|$,$d^{U}(A)$}, 
	col sep=comma, 
	column type=R{1cm}, 
	precision=2, 
	every head row/.style={before row=\hline\hline,after row=\hline}, 
	every last row/.style={after row=\hline\hline}, 
	columns/Operator/.style={ 
		string type, column type=l}, 
	columns/No./.style={ 
		string type, column type=L{1cm}}, 
	columns/Class/.style={ 
		string type, column type=R{1.5cm}}, 
	columns/$d(A)$/.style={column type=R{2cm}},
	columns/$d^{U}(A)$/.style={column type=R{2cm}},
	]{./Data/BraidsDF_Unitary_CloesestEquiv_TopforMaxBrL40.csv}
	\caption{Operators consisting of combinations of length 40, and their corresponding distance to the closest equivalence class in terms of the Makhlin invariants $g_1,g_2,g_3$, the norm of the $(1,1)$ element, and their unitary measure $d^U$.}
	\label{tab_closest_braids}
\end{table}

In this section thus far, we have focused on large combination lengths. Now, we reduce the length of the braid sequences while searching for $4 \times 4$ unitary operations that are close to some local equivalence classes. We perform an exhaustive search up to length 14, and remarkably, we can obtain a unitary matrix that is exactly within the local equivalence class $[\operatorname{SWAP}]$ using a braid of a length of only 9. Beyond $L=12$ we stop searching for $\operatorname{SWAP}$ gates as we have already achieved an optimal $\operatorname{SWAP}$ gate at $L=9$.
We summarize the results in Table \ref{tab_closest_small_braids}.

\begin{table}[htb!]
	\centering 	
	\begin{filecontents*}{./Data/TrialBraidsDF_ClosestEquiv_TopforMaxBrL14.csv}
Index,L,Operator,d,Class,Norm11,Unitarity
1,1,0,0.0,ID,1.0,0.0
2,2,00,0.0,ID,1.0,4.0e-16
3,3,000,0.0,ID,1.0,7.0e-16
4,4,0000,0.0,ID,1.0,4.0e-16
5,5,00000,0.0,ID,1.0,0.0
6,6,000000,0.0,ID,1.0,2.0e-16
7,7,0000000,0.0,ID,1.0,7.0e-16
8,8,00000000,0.0,ID,1.0,4.0e-16
9,9,234123012,0.0,SWAP,0.9999999999999994,4.7e-15
10,10,2341230120,0.0,SWAP,0.9999999999999994,4.6e-15
11,10,2223443222,0.4634536414463416,CNOT,0.9759111274880214,0.0475974712450606
12,11,23412301200,0.0,SWAP,0.9999999999999996,4.1e-15
13,11,22234430222,0.46345364144634,CNOT,0.9759111274880218,0.0475974712450608
14,12,423123041302,0.0,SWAP,0.9999999999999998,6.2e-15
15,12,223104403122,2.16908552831e-5,CNOT,0.975911127488021,0.0475974712450602
16,13,1133100000022,2.1690855283072532e-5,CNOT,0.9759111274880214,0.04759747124506086
17,14,03023010000322,2.1690855283079817e-5,CNOT,0.9759111274880216,0.047597471245057754
\end{filecontents*}
\pgfplotstabletypeset[
	columns={No.,$L$,Operator,$d(A)$,Class,$|M_{11}|$,$d^{U}(A)$}, 
	col sep=comma, 
	column type=R{1cm}, 
	precision=3, 
	every head row/.style={before row=\hline\hline, after row=\hline}, 
	every last row/.style={after row=\hline\hline}, 
	columns/Operator/.style={ 
		string type, column type=l}, 
	columns/No./.style={ 
		string type, column type=L{1cm}},
	columns/$L$/.style={ 
		string type, column type=L{1cm}},
	columns/Class/.style={ 
		string type, column type=R{1.4cm}}, 
	columns/$d(A)$/.style={column type=R{2cm}},
	columns/$d^{U}(A)$/.style={column type=R{2cm}},
	]{./Data/TrialBraidsDF_ClosestEquiv_TopforMaxBrL14.csv}
	\caption{Operators and their corresponding distance to the closest equivalence class in terms of the Makhlin invariants $g_1,g_2,g_3$, the norm of the $(1,1)$ element, and their unitary measure $d^U$.}
	\label{tab_closest_small_braids}
\end{table}


\pagebreak

\section{Conclusions}\label{sec_conclusions}

We numerically address the problem of topological compilation of two-qubit operations in the system of six Fibonacci anyons. This system gives rise to a five-dimensional Hilbert space which permits encoding of two qubits while leaving one-dimensional subspace as non-computational. This encoding has previously been studied by Cui et al. \cite{Cui_2Qubit_JoPA2019} who focused on the question of leakage-free two-qubit operations that are at the same time also entangling. They performed brute-force numerical searches for anyon braids with a length up to seven but found no entangling gates that would be exactly leakage-free. 

In this work, our objectives were different. Firstly, we focused on relaxing the compilation target from a single two-qubit operation, such as the $\operatorname{CNOT}$ gate, to its entire local equivalence class. This class is defined by the set of all two-qubit operations which are locally equivalent up to single-qubit transformations and it is uniquely determined by three numerical local invariants \cite{Makhlin_Invariants_QIP2002, ZhangValaBirgitta_GeometricTheory_PRA2003}.
This approach contrasts with finding a specific two-qubit operation, that is an element of the Lie group $SU(4)$ (or $U(4)$) which is characterized by 15 (or 16) parameters. Finding a specific operation is thus equivalent to finding a point in a fifteen-dimensional parameter space. In other words, using local invariants of two-qubit operations, we show that this problem, which is akin to finding a needle in a fifteen-dimensional haystack, can be considerably reduced. The resulting search, which then takes place in a rather intuitive three-dimensional space, has significantly improved convergence properties in both the length of the braids required as well as its accuracy compared to a search for a specific gate. 

Our numerical calculations were based on exhaustive search up to the braid length $L = 14$ and random brute-force search up to $L=40$. We point out that our search for the $\operatorname{CNOT}$ gate, using both the basic and extended alphabet, manifestly failed. None of the approximately leakage-free two-qubit operations could be considered close to the $\operatorname{CNOT}$ gate. On the other hand, our search for the local equivalence class $[\operatorname{CNOT}]$ led to satisfactory results. These are characterized by the unitarity measure $d^U(A)$ on the order of $10^{-2}$ and the distance to the local equivalence class $d^{\mathbf{E}}(A)$ on the order of $10^{-5}$ for the braids of the length $L=12$ and generally improving with the growing $L$. We do not observe a significant difference when using the basic or extended gate alphabets. However, more studies, perhaps with improved optimization techniques, may be required to fully understand this behavior. It seems plausible that the employed discrete gate alphabets, which effectively consist of only one two-qubit operation, may be too coarse to explore the relevant parameter space of two-qubit operations or their local equivalence classes more efficiently.

Finally, we study what local equivalence classes are naturally generated by our basic gate alphabet. We select braids that are unitary over the four-dimensional computational subspace and identify their local invariants. 
We observe the concentration of the resultant matrices along the line of so-called controlled unitary operations which connect the identity local equivalence class $[\ID]$ and the class $[\operatorname{CNOT}]$. We also observe a somewhat lesser concentration of the results along the line connecting the classes $[\operatorname{DCNOT}]$ and $[\operatorname{SWAP}]$. In fact, the class $[\operatorname{SWAP}]$ features prominently in our results as its generation is exactly leakage-free within standard numerical accuracy. On the other hand, all the entangling operations (and their local equivalence classes) are approximately leakage-free. This seems to be consistent with the results by Cui et al. \cite{Cui_2Qubit_JoPA2019} who suggested the negative for the open question of whether there are leakage-free entangling Fibonacci braiding gates.

\vspace*{-0.5cm}

\begin{section}{Aknowledgements} 
	P.C.B. acknowledges funding from Science Foundation Ireland through grant 21/RP-2TF/10019.
	C.A., J.M., and J.A. acknowledge the support of the Czech Science Foundation (23-0947S).
	J.V. acknowledges funding from Enterprise Ireland’s DTIF programme of the Department of Business, Enterprise, and Innovation, project QCoIr Quantum Computing in Ireland: A Software Platform for Multiple Qubit Technologies No. DT 2019 0090B.
	P.C.B. and J.V. acknowledge fruitful discussions with Paul Watts.
\end{section}
\newpage
\bibliographystyle{unsrt}
\normalem
\bibliography{TQC_bib}
\pagebreak

\end{document}